\documentclass[prc,aps,twocolumn,showpacs,showkeys]{revtex4}
\usepackage{graphicx}

\begin{document}

\newcommand{\p}{\partial} 
\newcommand{\ls}{\left(} 
\newcommand{\rs}{\right)} 
\newcommand{\beq}{\begin{equation}} 
\newcommand{\eeq}{\end{equation}} 
\newcommand{\beqa}{\begin{eqnarray}} 
\newcommand{\eeqa}{\end{eqnarray}} 
\newcommand{\bdm}{\begin{displaymath}} 
\newcommand{\edm}{\end{displaymath}} 

\title{Isospin emission and flow at 
high baryon density: a test of 
the symmetry potential}

\author{V. Giordano$^{1,2}$, M. Colonna$^{1}$, M. Di Toro$^{1,2,*}$, 
V.Greco $^{1,2}$, J. Rizzo$^{1,2}$}
\affiliation{$^{1}$ Laboratori Nazionali del Sud, INFN, Via S. Sofia 62,
  I-95123 Catania\\ 
$^{2}$ Physics and Astronomy Dept., University of Catania, Italy\\
$^{*}$ email: ditoro@lns.infn.it}

\begin{abstract}

High energy Heavy Ion Collisions ($HIC$) are studied in order to 
access nuclear matter
properties at high density. Particular attention is paid to
the selection of observables sensitive to the poorly known
symmetry energy at high baryon density, of large fundamental
interest, even for the astrophysics implications.
Using fully consistent transport simulations built on 
effective theories we test isospin observables
ranging from nucleon/cluster emissions to collective flows (in
particular the elliptic, squeeze out, part). 
The effects of the competition between stiffness and momentum dependence of
the Symmetry Potential on the reaction
dynamics are thoroughly analyzed. 
In this way we try to shed light on the controversial
neutron/proton effective mass splitting
at high baryon and isospin densities.
New, more exclusive, experiments are suggested.

\end{abstract}

\pacs{21.30.Fe, 21.65.-f, 25.75.Dw, 25.75.Ld}

\keywords{High Energy Heavy Ion Collisions, Fast Nucleon/Cluster Emission,
 Isospin Collective Flows, Symmetry Energy at High
Baryon Density, Symmetry Potentials, Neutron/Proton Effective Masses}

\maketitle

\section{Introduction}
In the 
nuclear Equation of State ($EoS$)
the symmetry energy $E_{sym}$ appears in the energy density  
$\epsilon(\rho,\rho_3) \equiv \epsilon(\rho)+\rho E_{sym} (\rho_3/\rho)^2
 + O(\rho_3/\rho)^4 +..$, expressed in terms of total ($\rho=\rho_p+\rho_n$)
 and isospin ($\rho_3=\rho_p-\rho_n$) densities. The symmetry term gets a
kinetic contribution directly from basic Pauli correlations and a potential
part from the highly controversial isospin dependence of the effective 
interactions \cite{baranPR,baoanPR,ditoroPPNP08}. Both at sub-saturation and 
supra-saturation
densities, predictions based on the existing many-body techniques diverge 
rather widely, see \cite{fuchswci}. 
We remind that the knowledge of the $EoS$ of 
asymmetric matter is very important at low densities (neutron skins,
nuclear structure at the drip lines, neutron distillation in fragmentation,
 neutron star formation and crust..) as well as at high densities (transition 
to a deconfined phase, neutron star mass/radius, cooling, hybrid structure,
 formation of black holes...). 
We take advantage of recent opportunities in 
theory (development of rather reliable microscopic transport codes for $HIC$)
 and in experiments (availability of very asymmetric radioactive beams, 
improved possibility of measuring event-by-event correlations) to present
results that could severely constrain the existing effective interaction 
models. We will discuss dissipative collisions at intermediate energies, i.e. 
in the range from $100AMeV$ up to  $1AGeV$. This will allow to probe
the symmetry term above normal density.
 
The reaction is simulated using transport codes based on effective 
mean field theories, with correlations/fluctuations included via hard 
nucleon-nucleon
collisions, see 
\cite{baranPR,chomazPR,rizzofluct}.  
The dynamical effects of a different density dependence of the isovector 
part of the $EoS$ are tested \cite{colonnaPRC57}.
Moreover a  particular attention is devoted to 
the isospin effects on the momentum dependence of the symmetry potentials,
i.e. to observables sensitive to a different neutron/proton effective mass 
in asymmetric matter.
The problem of Momentum Dependence in the Isovector
channel ($Iso-MD$) is still very controversial and it would be extremely
important to get more definite experimental information,
see the refs. 
\cite{baranPR,baoanPR,ditoroAIP05}. 
Exotic beams at intermediate energies are
of interest in order to have high momentum particles and to test regions
of high baryon (isoscalar) and isospin (isovector) density during the
reaction dynamics. 

The paper is arranged as follows. In Sect.2 the self-consistent transport 
approach to the reaction dynamics
is shortly described. Sect.3 is devoted to a detailed analysis of the used 
effective interactions, in particular of the isovector contributions to the 
local and non-local
part of the mean field potentials. The latter, which corresponds to the 
$Iso-MD$ case, leads to a splitting of the 
neutron-proton effective masses
in asymmetric matter. Results for $^{197}Au+^{197}Au$ are shown in Sect.4, 
selecting
the observables that would allow an independent study of the two 
isovector contributions.
Conclusions and perspectives are presented in Sect.5.

\section{The collision dynamics}

We perform {\it ab initio} collision simulations using
the microscopic Stochastic Mean Field (SMF) model.
It is based on mean field
 transport theory  with correlations included via hard nucleon-nucleon (NN)
collisions and  with inclusion of stochastic forces acting on the mean
phase-space trajectory
\cite{baranPR,guarneraPLB373,colonnaNPA642,fabbri04,chomazPR}.  Stochasticity 
is
essential in
order   to allow the growth of dynamical
instabilities leading to fragment production, as well as to obtain physical 
widths
of the observable distributions.  Moreover  it  will allow to perform
 event-by-event  correlation studies of great importance for  the very
complex reaction dynamics in this energy range.

The transport equation for the phase space distribution function, with the
Pauli blocking consistently evaluated, is integrated following a
 representation in terms of test particles of finite widths
\cite{guarneraPLB373,colonnaNPA642}. A detailed description of the procedure
is given in ref. \cite{baranPR}. Our code \cite{Alfio}
has been extended  by the introduction of momentum dependent mean fields
 (see next Section), which are 
 rather important in this energy range.  It has been also possible to
improve the numerical accuracy while even reducing the computing times
\cite{rizzoj_th}.

A parametrization of free nucleon-nucleon cross sections is used, with
isospin, energy and angular dependence  \cite{LiMachl94}. Low energy
NN collisions, mostly
forbidden
because of the Pauli blocking,  have  large cross sections and
 could  induce spurious effects in  the  presence of some numerical
inefficiency in the blocking procedure, due to the discretization of the
phase space.
In order to  avoid such problems  a cutoff value $\sigma_ {cut}=50mb$ is used
in our calculations.  A parallel ensemble method is employed in the 
implementation of the collision term. 

For   discussions of isospin dynamics  in this energy
regime it is essential  to have a reliable
procedure for fragment recognition, i.e. to  identify  free  and 
clusterized nucleons. 
A coalescence procedure is applied in an event by event analysis at the
``freeze-out'' time, i.e. when the resulting fragments are well separated 
in space and interacting only via Coulomb forces. 
The fragments are formed using a 
phase-space proximity criterion. Two particles are 
recognized to belong 
to the same cluster if they are sufficiently close in phase space. The 
procedure is applied to several random samplings of N 
(total number of nucleons) test particles from the full ensemble 
distribution at the freeze-out. The used 
coalescence parameters are $d_r=4.5fm$, $d_p=1.5fm^{-1}$, that reasonably 
reproduce the charge distributions in the same energy range \cite{santini05}.
We have also seen that the results do not show appreciable changes for 
small variations of the coalescence parameters. Since the random sampling 
usually is not exactly preserving energy and momentum conservation, we apply 
an additional constraint to impose them.

\section{Momentum dependence of the fields}

We adopt 
a  generalized form of the effective interactions, which can be easily
reduced to Skyrme-like forces, with momentum dependent terms  also  in the
isovector channel  \cite{rizzoj_th,ditoroAIP05,rizzoPRC72,rizzo08}.
The general structure of the  isoscalar and isovector Momentum Dependent (MD) 
 effective fields is
derived via an isospin asymmetric extension of the Gale-Bertsch-DasGupta  
(GBD)  force
 \cite{GBD,GalePRC41,BombaciNPA583,GrecoPRC59,BaoNPA735},  which  
corresponds to a Yukawian
 non-locality.

The energy density   is  parametrized
as follows:

\begin{equation}\label{energy}
 \varepsilon(\vec{r})=\varepsilon_{kin}(\vec{r})+\varepsilon_{A}
\left( \varrho(\vec{r})\right) +\varepsilon_{B}\left( \varrho(\vec{r})
\right) +\varepsilon_{C,z}\left(\vec{r})\right) 
\end{equation}

The kinetic term is:
\begin{equation}\label{ekin}
 \varepsilon_{kin}(\vec{r})=\int \frac{d^{3}p}{h^{3}}\left[ f_{n}
(\vec{r},\vec{p})+f_{p}(\vec{r},\vec{p})\right] \frac{p^{2}}{2m}
\end{equation}

The terms  $\varepsilon_{A}$ and $\varepsilon_{B}$ 
of Eq.(\ref{energy}) 
account for saturation properties, including the symmetry energy,

\begin{equation}\label{en_a}
\varepsilon_{A}(\vec{r})=\frac{{A}}{2}\frac{\varrho^{2}}
{\varrho_{0}}-\frac{{A}}{3}\left( \frac{1}{2}+x_{0}\right)\frac
{\varrho^{2}}{\varrho_{0}} \beta^{2} 
\end{equation}
\begin{equation}\label{en_b}
\varepsilon_{B}(\vec{r})=\frac{{B}}{\sigma+1}\frac{\varrho^
{\sigma+1}}{\varrho_{0}^{\sigma}}-\frac{2}{3}\frac{{B}}{\sigma +1}
\left( \frac{1}{2}+x_{3}\right)\frac{\varrho^{\sigma+1}}{\varrho_{0}^
{\sigma}} \beta^{2} 
\end{equation}
where $\rho$ is the "local" density, $\rho_0$ the saturation value and
$\beta = (N-Z)/(N+Z)$ represents the ``local'' asymmetry parameter.
The last term gives the momentum dependence and can be written as 
\cite{rizzo08}:

\begin{equation}\label{en_c}
\varepsilon_{C,z}(\vec{r})=\frac{8(C+2z)}{5\varrho_{0}}
I_{np}(\vec{r})+\frac{4(3C-4z)}{5\varrho_{0}}
(I_{nn}(\vec{r})+I_{pp}(\vec{r}))
\end{equation}

The momentum dependence is contained in the $\mathcal{I}_{\tau \tau'}$
terms, which are integrals of the form   
\begin{equation}
\mathcal{I}_{\tau \tau'}=\int d \vec{p} \; d \vec{p}\,'
f_{\tau}(\vec{r},\vec{p})
 f_{\tau'}(\vec{r},\vec{p}\,') g(\vec{p},\vec{p}\,') ~,
\label{momint}
\end{equation}
  with $\tau={P,N}$, for protons and neutrons.
Here  $f_{\tau}(\vec{r},\vec{p})$ are the nucleon phase space distributions
 for protons and neutrons
and the function  $g(\vec{p},\vec{p}\,') \equiv g[(\vec{p}-\vec{p}\,')^2]$
 determines the type of momentum dependence.
 A Skyrme-like momentum dependence is obtained  when we use the simple 
quadratic
form $g(\vec{p},\vec{p}\,') = (\vec{p}-\vec{p}\,')^2$.
A more general momentum dependence, in better agreement
with phenomenological optical potentials,
can be introduced by the function 
\cite{GBD,GalePRC41,BombaciNPA583,GrecoPRC59,BaoNPA735}
\begin{equation}
g(\vec{p},\vec{p}\,')=
\left[
1+ \left( \frac{\vec{p}-\vec{p}\,'}{\Lambda}\right)^2
\right]^{-1} . 
\label{momform}
\end{equation}

For symmetric nuclear matter ($\beta=0$) the energy density Eqs.
(\ref{ekin},\ref{en_a},\ref{en_b},\ref{en_c}) reduces to the 
parametrization proposed by
Gale, Bertsch and Das Gupta \cite{GBD,GalePRC41}, that for low densities 
and momenta gives the
same results of the Skyrme-Gogny force, see 
\cite{GrecoPRC59}. 

The EoS of symmetric matter is fixed by the parameters A, B, C, $\sigma$ 
and $\Lambda$. We choose
the values $A=-111.3MeV$, $B=141.3MeV$, $C=-64.5MeV$, $\sigma=7/6$ and 
$\Lambda=1.5p_F^0$ (with
$p_F^0$ Fermi momentum at normal density), which provide a rather 
Soft EoS with compressibility
$K\simeq 215MeV$ and an isoscalar effective mass $m^*/m=0.67$ at 
saturation density
$\rho_0=0.16fm^{-3}$. We note that an indication for a Soft Eos 
(Isoscalar) even at
higher densities comes also from heavy ion data (collective flows and 
meson production)
at intermediate energies \cite{dan02}.

\begin{figure}
\unitlength1cm
\begin{center}
\includegraphics[width=6.0cm]{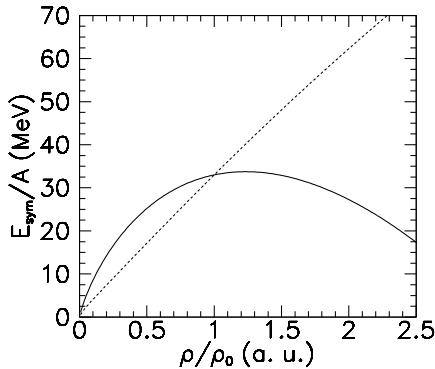}
\caption{Density dependence of the  symmetry energies  used in the simulations
presented here:  asy-soft (solid) and asy-stiff (dashed).
}
\end{center}
\label{esymdens}
\end{figure}

From the energy density one derives the mean field potentials as
$U_{\tau}(\vec{r},\vec{p})=\delta\varepsilon/\delta f_{\tau}$. Thus the above
energy density implies a momentum dependent mean field interaction.
The momentum dependence is isoscalar if, in the Eq.(\ref{en_c}),
the coefficients 
in front of
$I_{np}(\vec{r})$ and in front of the combination   
$(I_{nn}(\vec{r})+I_{pp}(\vec{r}))$ are equal 
(corresponding to the $C=8z$ case),
 but it can also get an isovector part, if they are different.

The isovector momentum dependence implies different effective masses
for protons and neutrons given as
$\frac{m^*_{\tau}}{m}=(1+\frac{m}{\hbar^2 p}\frac{\partial U_{\tau}}
{\partial p})^{-1}$, for $ p=p_{F,\tau}$, at fixed density. 

\begin{figure}[htb]
\centering
\vskip 0.7cm
\includegraphics[width=8.0cm]{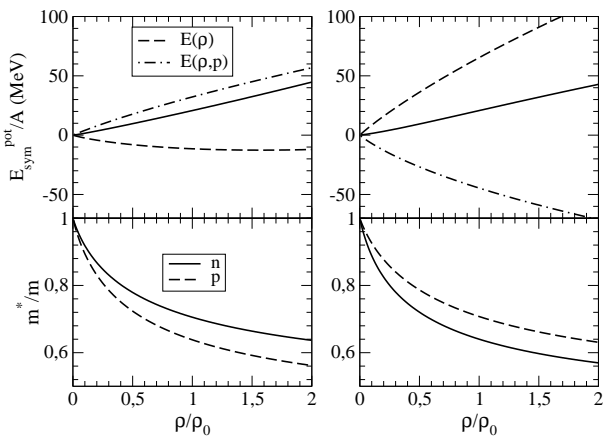}
\caption{Upper Panels: Density dependence of the potential symmetry energy 
(Solid Lines), in 
the $Asystiff$
choice. Dashed lines 
refer to
local contributions, dot-dash lines to momentum-dependent ones, see text. 
Left: 
 $m^*_n>m^*_p$ parametrization; Right: $m^*_n<m^*_p$ case.
Lower Panels: corresponding behavior of neutron/proton effective masses as a 
function 
of the density,
for an asymmetry $\beta=0.2$.}
\label{esymastar} 
\end{figure}

Here we probe the sensitivity of isospin observables
to two essentially different density dependence of the symmetry energy around 
saturation:
{\it Asy-soft} with a smooth behavior below saturation, 
 and even a decrease at high densities, vs.
{\it Asy-stiff} which always shows instead a rapid increase, roughly
proportional to the density \cite{baranPR,colonnaPRC57,ditoroPPNP08}.
In fact the two choices are well characterized by the slope parameter 
$$L\equiv 3\rho_0(dE_{sym}/d\rho)_{\rho=\rho_0}$$
which is of the order of $10-20$MeV in the Asysoft case, just from 
the kinetic contribution
to the symmetry term, and around $70-100$MeV in the Asystiff 
parametrization with also a repulsive 
potential part  
\cite{colonnaPRC57,baranPR}.
In Fig.1 we show the density dependence for these two typical 
choices. 

\begin{center}
{\large \bf Table 1.}~Parameters $x_0$, $x_3$,$z$ for opposite 
mass splittings, with
the same $E_{sym}=33MeV$,
 for $Asysoft$ (L=19MeV) and $Asystiff$ (L=95MeV).
\begin{tabular}{c|c|c|c|c|c|c|c} \hline
 Asysoft &    &     &      \\ \hline
 Mass splitting & $x_0$ & $x_3$ & $z(MeV)$ \\ \hline
$m_n^*>m_p^*$ & 1.111  & 1.196   & -35.467  \\ 
$m_n^*<m_p^*$ & 5.212  & 2.988   & 44.971 \\ 
$m_n^*=m_p^*$ & 3.165  & 2.094   & 4.811 \\ \hline
Asystiff &     &     &     \\ \hline
Mass splitting & $x_0$ & $x_3$ & $z$ \\ \hline
 $m_n^*>m_p^*$ & -1.614  & -1.210   & -35.467  \\
  $m_n^*<m_p^*$ & 2.487  & 0.582   & 44.971   \\ 
$m_n^*=m_p^*$ & 0.440  & 0.312   & 4.811 \\ \hline
\end{tabular}
\end{center}

When we use momentum-dependent interactions we have also contributions 
to the symmetry energy 
from the non-local terms. 
The presence 
of an interplay 
between the $C$ and $z$ parameters in our form of the effective 
interaction allows
an independent study of the dynamical effects of the stiffness of 
the symmetry term and of the 
neutron/proton effective mass splitting. This can be easily done just 
varying the 
$x_0$, $x_3$ and $z$ parameters. In Table I we report the used set 
of parameters. The 
symmetry energy
at saturation is always fixed to $E_{sym}(\rho_0)=33MeV$. 

\begin{figure}[h!]
\centering
\includegraphics[width=8.0cm,angle=0]{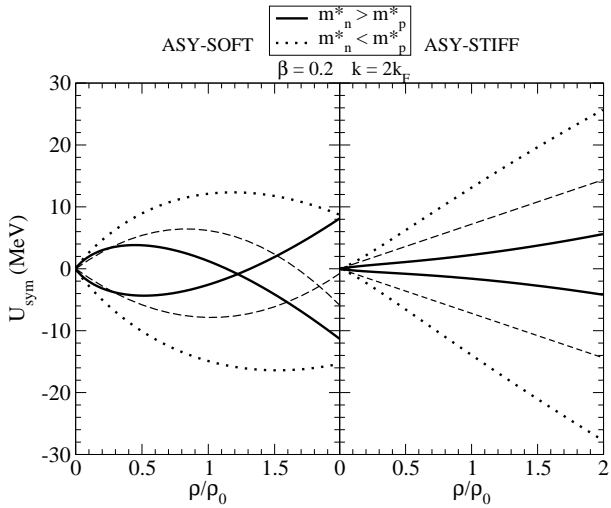}
\caption{Density dependence of neutron(upper)-proton(lower) potentials for
an asymmetry $\beta=0.2$ for the $Asysoft$(left) and $Asystiff$(right) 
choices. 
Dashed: No momentum dependence. Momentum dependent potentials at $k=2k_F$:
solid lines for the $m_{n}^{*} > m_{p}^{*}$ case, dotted lines for the opposite
$m_{n}^{*} < m_{p}^{*}$ choice.}
\label{Upot}
\end{figure}

In Fig.\ref{esymastar} we plot the density 
dependence of the potential part 
of the symmetry energy, 
in the Asystiff case,
for the two choices of the n/p mass splitting (solid lines, upper panels). 
We also separately report the 
contributions from 
the momentum-dependent, $E(\rho,p)$, and the density dependent, $E(\rho)$, 
part 
of the EoS, whose sum gives the total $E_{sym}^{pot}$. 
A change in the sign of the mass
splitting is related to opposite behaviors of these two contributions, 
exactly like it happens in
Skyrme-like forces, see sections (2.1-2.2) of ref.\cite{baranPR}. 
The lower panels show
the density dependence of the corresponding mass splitting, for an
asymmetry parameter $\beta=0.2$ (the $^{197}$Au asymmetry). In order to
probe the mass splitting effects on the heavy ion dynamics we have chosen
parametrizations that give almost opposite splittings at all densities.

In Fig.\ref{Upot} we present the density dependence of the neutron/proton 
symmetry potentials,
 for the two stiffness of the symmetry term, evaluated in the 
case without momentum 
dependence (dashed lines) and in the momentum dependence ($Iso-MD$) case 
for the  
$m_{n}^{*} < m_{p}^{*}$ 
(dotted) and the opposite $m_{n}^{*} > m_{p}^{*}$ (solid) choices.  We see 
that the momentum dependence
modifies the effect of the symmetry term stiffness on the nucleon 
potentials, with differences that
become more appreciable with increasing nucleon momenta. From this 
figure we can already
predict large effects of the effective mass splitting at high momenta.

\begin{figure}[h!]
\centering
\includegraphics[width=8.0cm,angle=0]{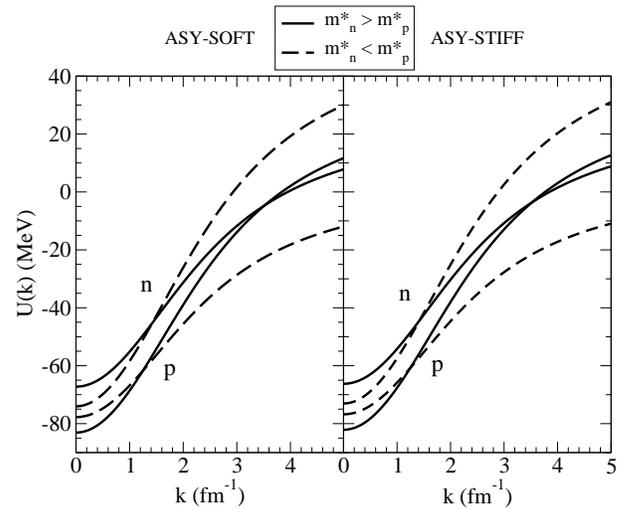}
\caption{Momentum dependence of neutron-proton potentials at saturation 
density and asymmetry $\beta=0.2$, for the two splitting choices 
$m_{n}^{*}<m_{p}^{*}$ (dashed) and 
$m_{n}^{*}>m_{p}^{*}$ (solid). Left panel: $Asysoft$ $Iso-Eos$. 
 Right panel: $Asystiff$ case}
\label{mdpot}
\end{figure}

This is shown more explicitly in Fig.\ref{mdpot} where we see the 
momentum dependence
of the neutron-proton potentials at saturation density for the two mass
splitting choices,
always for a "typical" $\beta=0.2$ asymmetry ($^{124}Sn, ^{197}Au$...).
The plot is for the $Asysoft$ (left panel) and the $Asystiff$ (right) 
symmetry term, and in fact it is not much different. Indeed
we can see also from the previous Fig.\ref{Upot}  
that at normal
density the difference between neutron and proton 
potentials is
almost the same for the two asy-stiffness, even in the case of 
$Iso-MD$ interactions..

The Figs.\ref{Upot} and \ref{mdpot} suggest the presence of 
interesting Isoscalar and 
Isovector $MD$ effects on the reaction dynamics: 
\begin{itemize}
\item 
{$Isoscalar$. In general the momentum dependence gives more 
attractive potentials at low 
momenta, $p<p_F$, and more repulsive at high $p$, $p>p_F$. 
In the reaction dynamics we expect
the more energetic nucleons to be fast emitted and to suffer 
less collisions \cite{GBD}. As a 
consequence we will 
have less stopping of the matter and less compression. 
The isoscalar EoS becomes stiffer
 when the momentum dependence is included.}
\item
{$Isovector$. Isospin effects on the momentum dependence imply 
different slopes around $p_F$
for neutrons and protons, as clearly shown in Fig.\ref{mdpot}, and so 
the larger repulsion
above $p_F$ is different. In the case $m_{n}^{*}<m_{p}^{*}$ the high 
momentum neutrons will
see a more repulsive field with respect to the high-$p$ protons. The 
opposite will happen
in the $m_{n}^{*}>m_{p}^{*}$ case. The fast nucleon emission will be 
directly affected: in the
$m_{n}^{*}<m_{p}^{*}$ case we expect a larger $n/p$ yield for nucleons 
emitted in central collisions
and a larger neutron $Squeeze-out$ (elliptic flow) in semicentral 
collisions in heavy ion reactions
at intermediate energies, in particular for high $p_t$ (transverse momentum) 
selections. 
In fact in the interacting, high density, early stage of the reaction dynamics
the pressure is built from
violent nucleon-nucleon collisions and the high $p_t$ particles will carry
the maximal information on high density and momentum dependence of the 
symmetry potentials. The azimuthal distributions (elliptic flows) will be
particularly affected since particles mostly retain their high transverse
momenta escaping along directions orthogonal to the reaction plane without
suffering much rescattering processes. 

We will test those 
predictions
also for n-rich vs. n-poor light ions, like ($^3H$, $^3He$), easier to 
detect. Since, as already 
noted, the symmetry potentials are not very different in the 
Asystiff/Asysoft choice for density 
range probed at intermediate energies
 (see the discussion in the next Section) , we can expect that 
the $Mass-Splitting$
effect could be even larger than the one related to the different 
stiffness of the symmetry term. 

A final interesting point is about the crossing of the two predictions, 
with opposite mass-splitting
choice, at high momenta, see Fig.\ref{mdpot}. Roughly this should 
happen around the Fermi 
momentum corresponding to
the density reached in the compressed source of the fast emitted nucleons. 
This represents an
independent way to check the maximum density reached during the collision.
}
\end{itemize}

\section{Results on $^{197}Au$+$^{197}Au$ Reactions at Intermediate Energies}

Our Stochastic Mean Field ($SMF$)
 transport code has been implemented with 
 $Iso-MD$  symmetry potentials,
with a different $(n,p)$ momentum dependence, as discussed in detail
in the previous section. This will allow to follow the dynamical
effect of opposite n/p effective mass splitting while keeping the
same density dependence of the symmetry energy   \cite{rizzo08}.

We present here some results for  $^{197}Au+^{197}Au$
 reactions at $400AMeV$ and  $600AMeV$ \cite{vale08}. For central 
collisions in the interacting zone we can reach baryon densities about
$1.7-1.8 \rho_0$ in a transient time of the order of 15-20 fm/c. The system 
is quickly expanding and the Freeze-Out time is around 50fm/c. At this time
we have a dominant Coulomb interaction among the reaction products. All the 
results presented here refer to this time step. Secondary decays of 
excited primary
fragments are not accounted for. In fact this will not affect too much the 
properties of nucleons and light ions at high transverse momenta 
mostly discussed 
in this work.

\begin{figure}
\centering
\includegraphics[width=8.0cm]{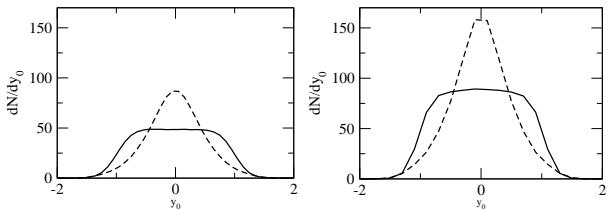}
\caption{197Au+197Au at 400AMeV (left) and 600AMeV (right), central collision.
 Longitudinal (solid lines) and Transversal (dashed lines) scaled 
rapidity distributions
 of Z=1-3 ions.
}
\label{dndy}
\end{figure}

\begin{figure}
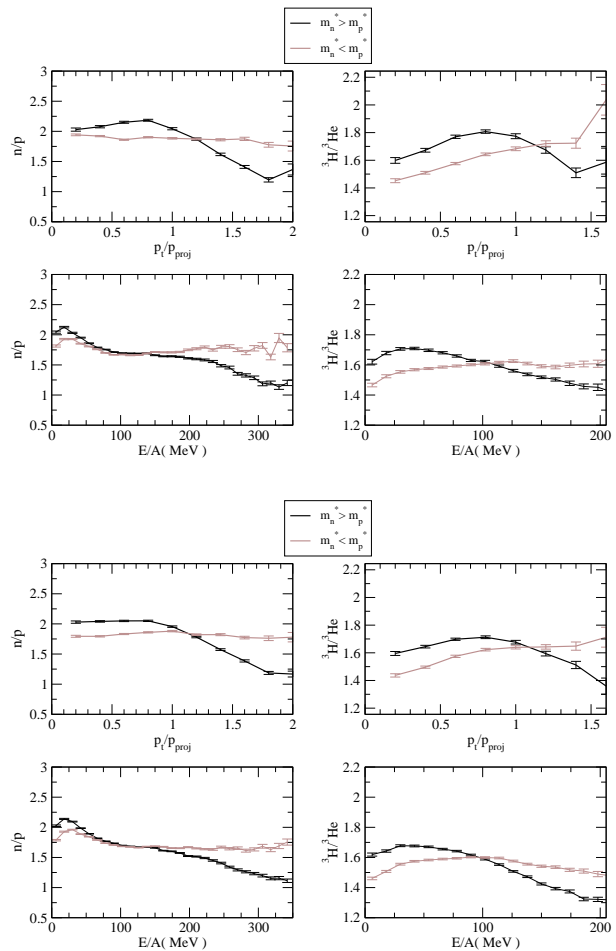

\centering
\includegraphics[width=8.0cm]{val6a.eps}
\vskip 0.5cm
\includegraphics[width=8.0cm]{val6b.eps}
\caption{197Au+197Au at 400AMeV, central collision.Isospin content of 
nucleon (left) and light ion (right) emissions vs. $p_t$ 
at midrapidity, $\mid y_0 \mid <0.3$, (upper) and
 kinetic energy (lower), for all rapidities,
for the two nucleon mass splitting choices.
Top Panels:
Asysoft; Bottom Panels: Asystiff.}
\label{ratios400}
\end{figure}

\begin{figure}
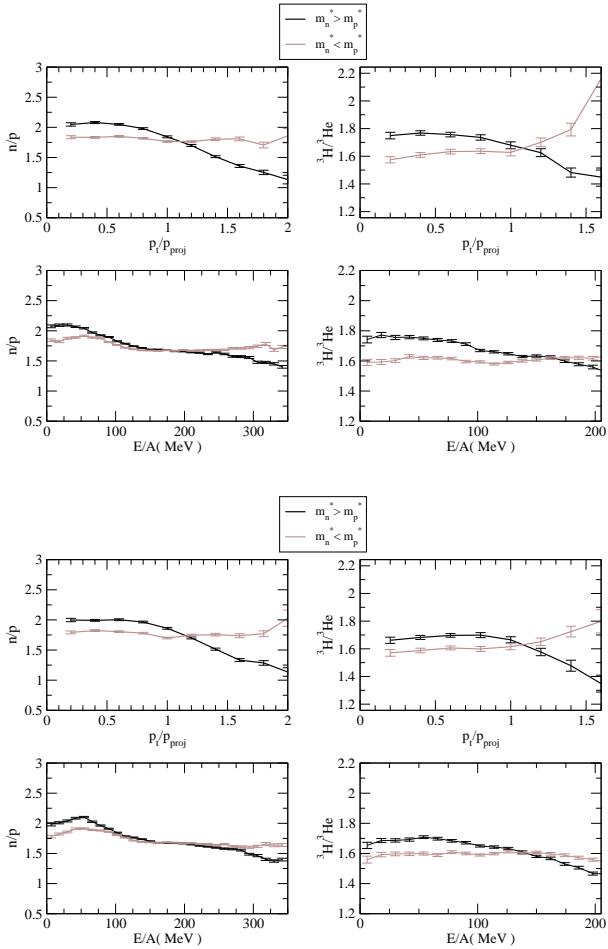

\centering
\includegraphics[width=8.0cm]{val7a.eps}
\vskip 0.5cm
\includegraphics[width=8.0cm]{val7b.eps}
\caption{Same as in Fig.(\ref{ratios400}) for 197Au+197Au at 600AMeV, 
central collision.}
\label{ratios600}
\end{figure}

For each beam energy and centrality we run 50 independent events.
In order to show how global properties of the reaction are reproduced
 by the simulations, 
 in Fig.\ref{dndy} we report the longitudinal and transversal rapidity 
distributions of protons and light ions (Z=1-3) in central collisions 
at the two beam 
energies. We use the $CM$ scaled rapidity $y_0 \equiv y_{cm}/y_{proj}$, 
vs. the projectile
rapidity. A good parameter to evaluate the stopping power of the collision 
is the
$Vartl$ quantity, i.e. the ratio between the variances of the transverse and 
longitudinal
rapidity distributions, recently suggested by the FOPI collaboration, 
\cite{reis06,reis09}.
In our calculation we get values around 0.75, not much different at the 
two beam energies,
comparable to data although no experimental trigger is included. We note 
that this estimation
of the stopping power appears not dependent on the isovector contributions 
to the EoS. 
 This clearly shows the difficulty of the search for well measurable 
isospin effects in 
the reaction
 dynamics. Here we will focus our attention on the isospin content of fast 
emitted nucleons 
and light ions and on isospin collective flows.

\subsection*{Isospin ratios of fast emitted particles}

In Figs.\ref{ratios400},\ref{ratios600} we present the  $(n/p)$ and
 $^3H/^3He$ yield 
ratios at freeze-out,
for two choices of Asy-stiffness and Mass-splitting, vs. transverse
momentum in a mid-rapidity selection (upper curves) and kinetic energy (
all rapidities, lower curves). In this way we can
separate particle emissions from sources at different densities, as discussed
in the previous Section.

For both beam energies we clearly observe the opposite effect of the 
different mass 
splitting in the low and high
momentum regions, as expected from Fig.\ref{mdpot}. 
E.g. in the $m_{n}^{*}<m_{p}^{*}$ case the neutrons see 
a less
repulsive potential at low momenta and a more repulsive one at high $p_t$. 
The curves
in the opposite mass-splitting show exactly the opposite behavior.
We note some interesting features: 

\noindent
i) The effect is almost not dependent
on the stiffness of the symmetry term. At high $p_t$, where particles
mostly come from high density regions, the larger repulsion seen by
neutrons in the Asy-stiff case, leading to an enhanced emission, is 
compensated by the larger Coulomb repulsion in the remaining matter, 
favoring proton emission. On the other hand, at low $p_t$, the sensitivity
to the Asy-stiffness is lost due to the mixing of sources at different 
densities, also for central rapidities, during the radial expansion. 

 \noindent
ii) The curves are crossing at
$p_t \simeq p_{projectile}= 2.13 fm^{-1}$. The crossing nicely corresponds
to the Fermi momentum of a source at baryon density $\rho \simeq 1.6 \rho_0$, 
 \cite{ditoroAIP05,rizzoPRC72,rizzo08}. 

\noindent
iii) The results appear not very sensitive to the beam energy going from $400$
to $600$ AMeV, likely because the reached maximum density is not much different
in the two cases. The same is observed for the isospin flows analyzed in the
following Subsection.

We remark that all the effects discussed before should be also present for the
$^3H/^3He$ yield ratios, more easily detected. Particularly interesting is the
predicted large increase at high $p_t$ in the $m_{n}^{*}<m_{p}^{*}$ choice. 
Some preliminary FOPI results seem to indicate this trend \cite{reis09}, 
but more data
are needed. It is encouraging that we already see a good $Iso-MD$ 
dependence of rather inclusive nucleon/cluster emission data. In presence of a 
good statistics for the detection of high $p_t$ particles, a further selection
at high azimuthal angles and central rapidities would certainly enhance the
sensitivity to the momentum dependence of the Symmetry Potentials. This
will introduce the discussion of isospin elliptic flows of the following
Subsection.

\begin{figure}[t]
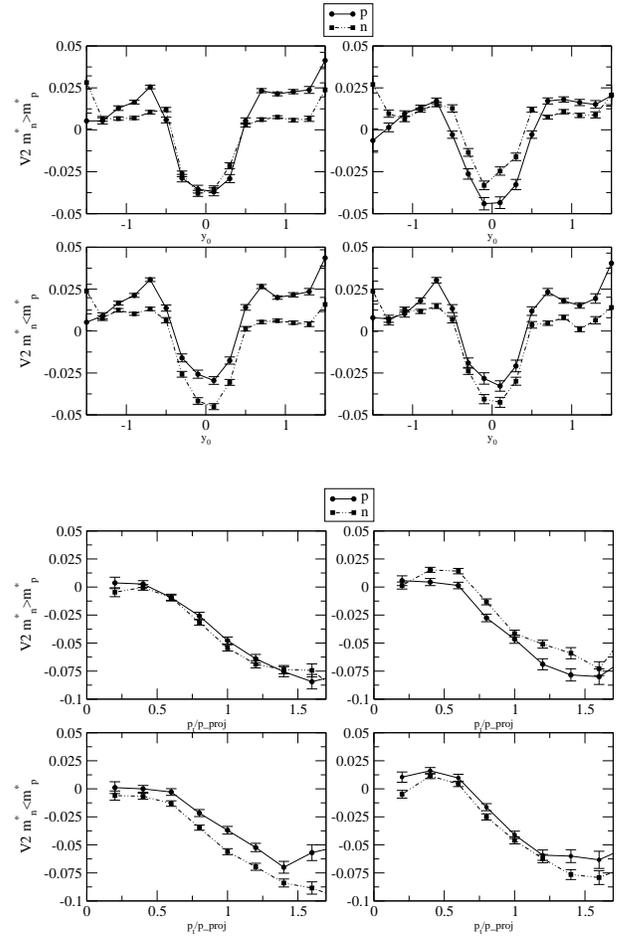

\centering
\includegraphics[width=8.0cm]{val8a.eps}
\vskip 0.5cm
\includegraphics[width=8.0cm]{val8b.eps}
\caption{
Proton (thick) and neutron (thin) $V_2$ flows in a 
semi-central
reaction Au+Au at 400AMeV.
Top Panels:
Rapidity dependence. 
Bottom Panels:
Transverse momentum dependence at midrapidity, $\mid y_0 \mid <0.3$.
Upper curves for $m_n^*>m_p^*$, lower curves for
the opposite splitting $m_n^*<m_p^*$. Left: Asystiff. Right: 
Asysoft.}
\label{v2ypt400}
\end{figure}

\begin{figure}[t]
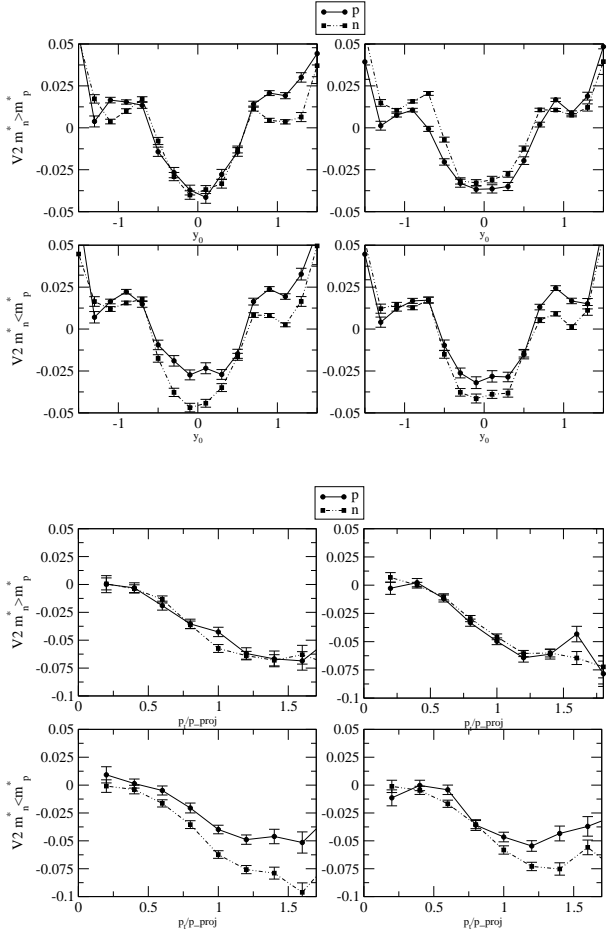

\centering
\includegraphics[width=8.0cm]{val9a.eps}
\vskip 0.5cm
\includegraphics[width=8.0cm]{val9b.eps}
\caption{
Same as for Fig.\ref{v2ypt400} for the
semi-central
reaction Au+Au at 600AMeV.}
\label{v2ypt600}
\vskip 0.5cm
\end{figure}

\subsection*{Isospin flows}

Collective flows are very good candidates since they are 
expected to be 
very sensitive to the momentum
dependence of the mean field, see \cite{DanielNPA673,baranPR}.
The flow observables can be seen
respectively as the
first and second coefficients from the Fourier expansion of the
azimuthal distribution:
$\frac{dN}{d\phi}(y,p_t)=N_0[1+V_1cos(\phi)+2V_2cos(2\phi)]$, 
where $p_t=\sqrt{p_x^2+p_y^2}$ is the transverse momentum and $y$
the rapidity along beam direction. 
The transverse flow, 
$V_1(y,p_t)=\langle \frac{p_x}{p_t} \rangle$,
provides information on the anisotropy of 
nucleon emission on the reaction plane.
Very important for the reaction dynamics is the elliptic
flow,
$V_2(y,p_t)=\langle \frac{p_x^2-p_y^2}{p_t^2} \rangle$.
 The sign of $V_2$ indicates the azimuthal anisotropy of emission:
on the reaction
plane ($V_2>0$) or out-of-plane ($squeeze-out,~V_2<0$)
\cite{DanielNPA673}.

\begin{figure}[t]
\centering
\includegraphics[width=7.5cm]{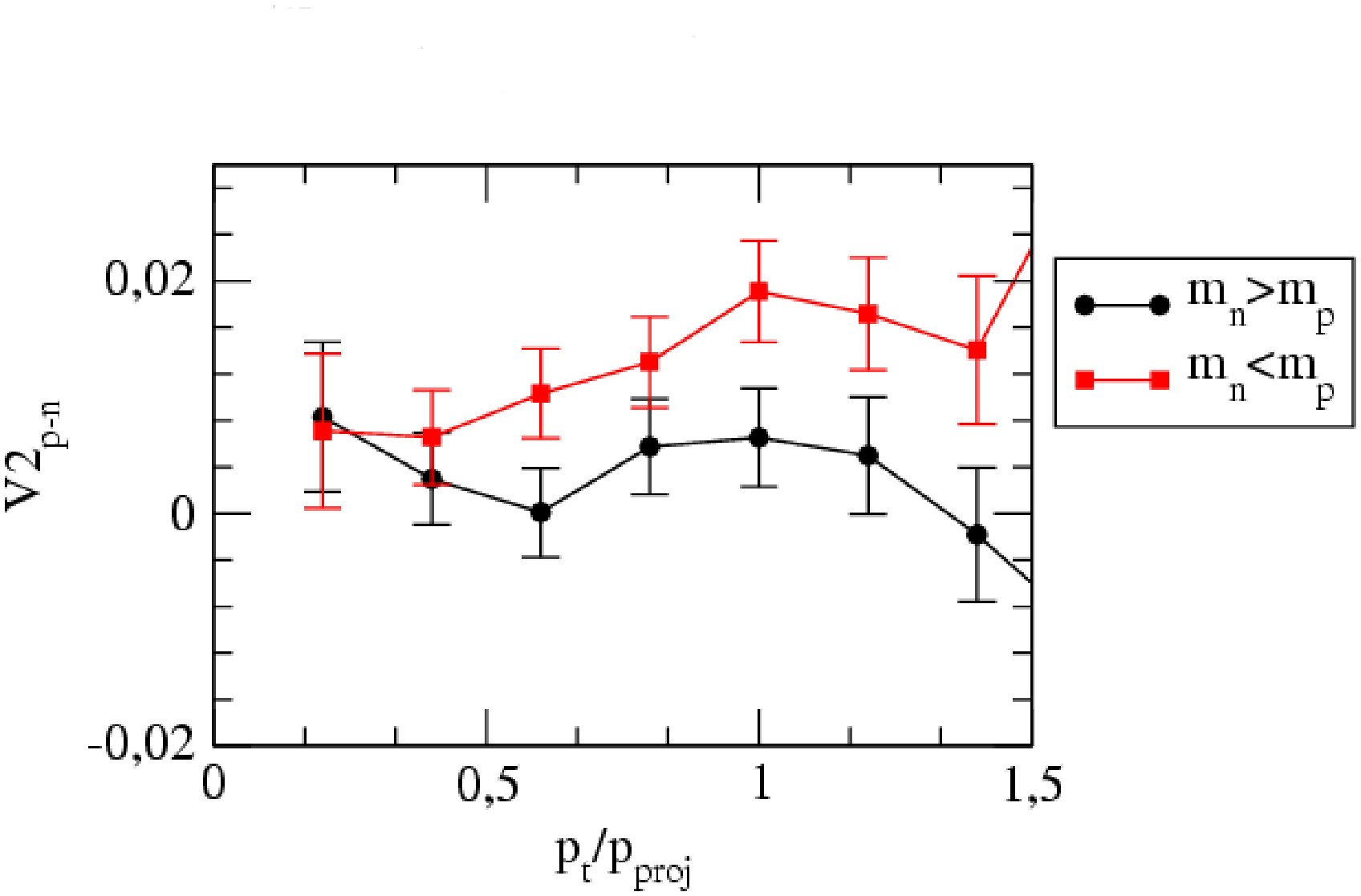}
\caption{
Transverse momentum dependence of the difference between proton 
and neutron $V_2$ flows, at mid-rapidity, $\mid y_0 \mid < 0.3$, in a 
semi-central
reaction Au+Au at 400AMeV. Asystiff choice.} 
\label{v2dif}
\end{figure}

\begin{figure}[t]
\centering
\includegraphics[width=8.0cm]{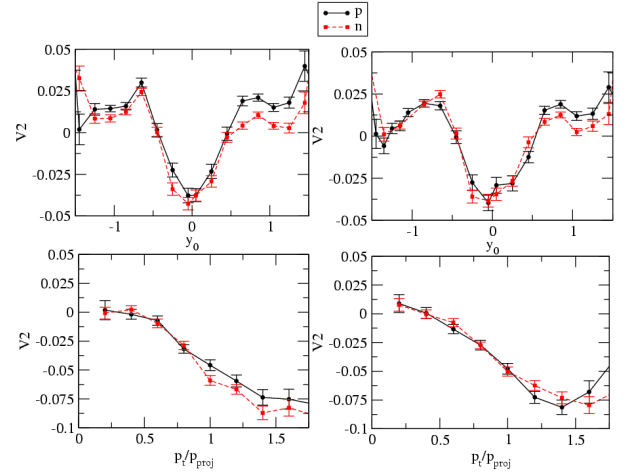}
\caption{
Proton (thick) and neutron (thin) $V_2$ flows in a 
semi-central
reaction Au+Au at 400AMeV, for equal (n.p)-effective masses.
Top Panels:
Rapidity dependence. 
Bottom Panels:
Transverse momentum dependence at midrapidity.
Left: Asystiff. Right: 
Asysoft.}
\label{nosplit}
\end{figure}

Isospin effects on collective flows have been studied within the $UrQMD$ 
transport model
in order to probe the influence of the symmetry repulsion at high densities
 \cite{qli0506,qli06}, here we focus the attention on the mass-splitting 
contributions.
For the same Au+Au reactions, in a semicentral selection, we present in Figs.
\ref{v2ypt400}, \ref{v2ypt600} (Upper Panels) the rapidity dependence of 
$(n/p)$ $V_2$ for 
different choices of the Asy-stiffness and effective mass splitting. We 
observe the relevance of the latter: at mid-rapidity the neutron 
squeeze-out is much 
larger in the 
$m^*_n < m^*_p$ case independently of the stiffness of the symmetry term. 
We note however that in the Asysoft case we see an inversion of the 
neutron/proton squeeze-out
at mid-rapidity for the two effective mass-splittings. Good data 
seem to be suitable
to disentangle $Iso-MD$ potentials.

The mass-splitting effect is large at high $p_t$ (Bottom Panels), again in a 
mid-rapidity selection, 
as expected for particle emitted from higher density regions.
Here the results are also slightly depending on the Asy-stiffness,
 with large neutron $squeeze-out$ effects in the Asystiff case.
 

A good 
sensitive observable seems to be the 
${\it Difference}$ of $(p/n)$ elliptic flows, $V_2(p)-V_2(n)$, 
shown in Fig.\ref{v2dif}
vs. transverse momentum at mid-rapidity, that is systematically larger
in the  $m^*_n < m^*_p$ case.

In order to have a clear idea of the relevance of the (n,p) mass 
splitting on the 
fast nucleon emissions we present in Fig.\ref{nosplit} the neutron/proton 
elliptic flows
for semicentral $Au+Au$ collisions at $400AMeV$ evaluated 
with the parametrizations giving $m^*_n=m^*_p$ 
for the $^{197}Au$ asymmetry $\beta \simeq 0.2$ \cite{equalmass}.
Now the isospin 
effects are only related to the different stiffness of the symmetry 
term at suprasaturation
density. We see that, at variance with the mass-splitting results of 
Fig.\ref{v2ypt400}, the
rapidity distributions (top panels) are not much affected, with a 
slightly larger neutron 
squeeze-out
in the Asystiff case. Consistently we see some difference in the 
transverse momentum dependence
at mid-rapidity (bottom panels) only at very large $p_t$.

\begin{figure}[t]
\centering
\includegraphics[width=7.0cm]{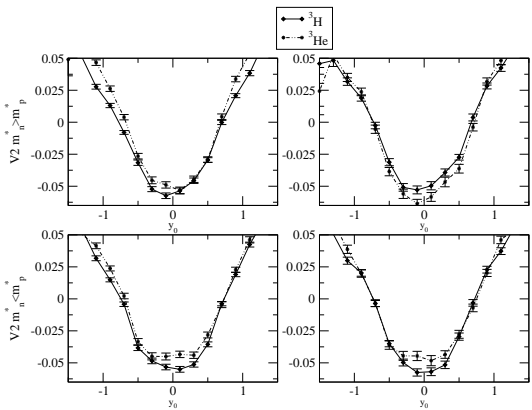}
\caption{
Rapidity dependence of Triton (thin)
and $^3He$ (thick) $V_2$ flows in a 
semi-central
reaction Au+Au at 400AMeV. Upper curves for $m_n^*>m_p^*$, lower curves for
the opposite splitting $m_n^*<m_p^*$. Left Panels: Asystiff. Right Panels: 
Asysoft.}
\label{tvshe400}
\end{figure}

\begin{figure}[t]
\centering
\includegraphics[width=7.0cm]{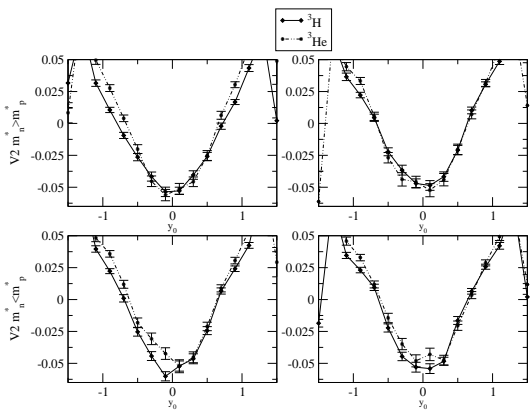}
\caption{
Same as in Fig.\ref{tvshe400} for a
semi-central
reaction Au+Au at 600AMeV.}
\label{tvshe600}
\end{figure}

\begin{figure}[t]
\centering
\includegraphics[width=8.0cm]{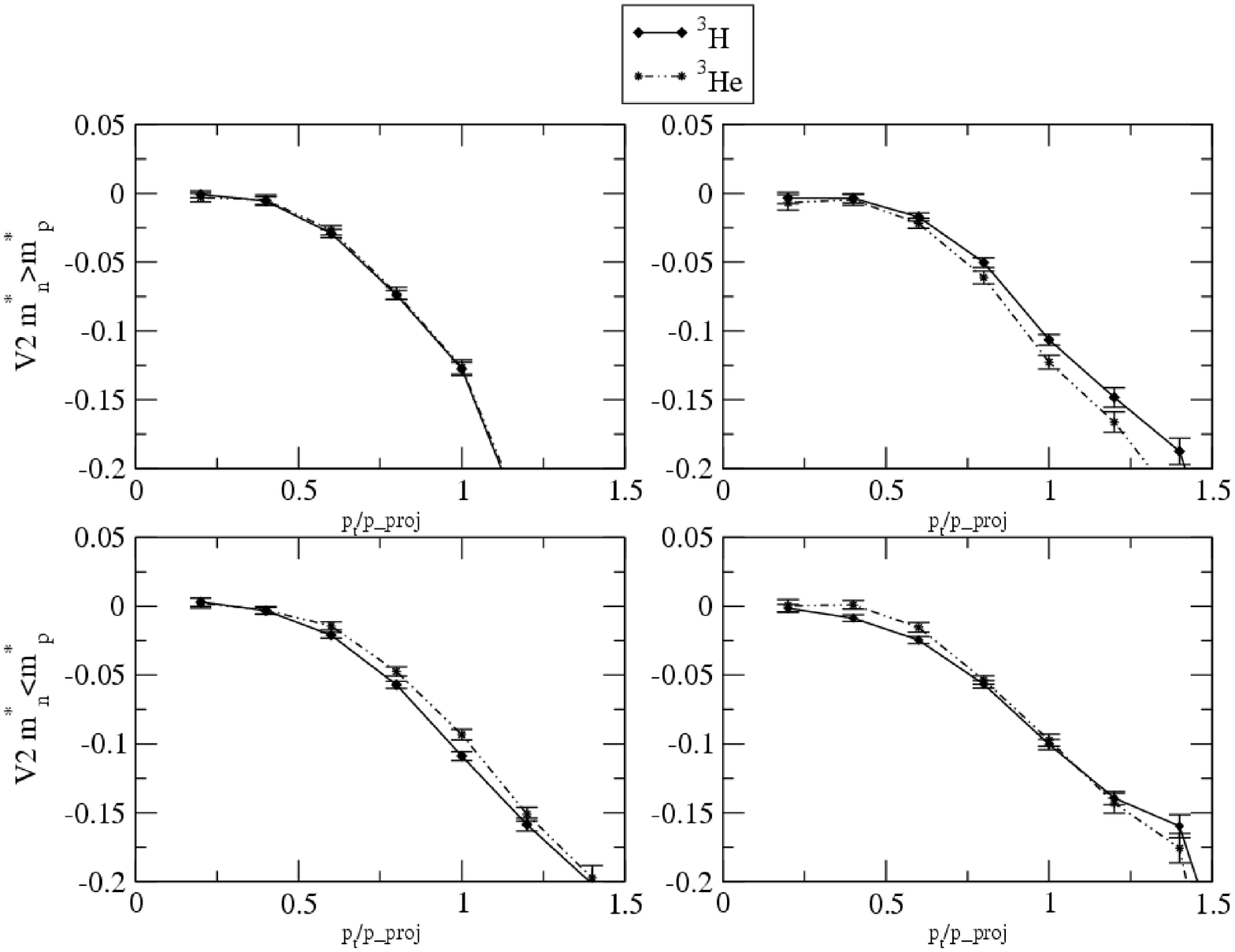}
\caption{Transverse momentum dependence at mid-rapidity
of $^3He$ (thick) and $^3H$ (thin) $V_2$ flows in a 
semi-central
reaction Au+Au at 400AMeV.
Upper curves for $m_n^*>m_p^*$, lower curves for
the opposite splitting $m_n^*<m_p^*$. Left: Asystiff. Right: 
Asysoft.}
\label{teliov2pt}
\end{figure}

Due to the difficulties in
measuring neutrons, we have analyzed the isospin sensitivity  
of light isobar flows, like $^3H$ vs. $^3He$ and so on.
We still see effective mass splitting effects, although 
slightly reduced,
 as shown in Figs.
\ref{tvshe400}, \ref{tvshe600} for the triton/$^3He$ case. 

As in the nucleon elliptic flow, at mid-rapidity the triton squeeze-out is 
larger in the 
$m^*_n < m^*_p$ case independently of the stiffness of the symmetry term. 
Again in the Asysoft case we see an inversion of the  $^3H$ vs. $^3He$  
squeeze-out
at mid-rapidity for the two choices of the mass-splitting. 

In Fig.\ref{teliov2pt} we show also the the transverse momentum dependence
of the $^3He,^3H$ elliptic flows at mid-rapidity ($\mid y_0 \mid < 0.3$)
at $400AMeV$. Some mass-splitting effects can be seen at high $p_t$.
This should be well observed in the flow difference. Unfortunately 
with the present number of events we cannot present a figure (like the
Fig.\ref{v2dif} of the $(p,n)$ case) since we have too large error bars
for the lack of statistics. 

We note that increasing the beam energy from $400$ to $600$ AMeV the symmetry
potential effects are not much changing, as already seen also for the isospin 
ratios of fast emitted particles discussed in the previous Subsection.
In this respect an interesting positive result is coming from
preliminary FOPI data on $^3H$ vs. $^3He$ flows, for Au-Au collisions
at beam energies
extended up to 1.5AGeV. The triton $V_2$ shows 
a larger squeeze-out
at mid-rapidity (in a relatively high transverse momentum selection)
 \cite{reis09}, clearly increasing with the beam energy, 
consistent with the reached higher baryon 
densities.

\section{Conclusion and perspectives}
The paper has been mainly devoted to the study of the effects of the 
momentum dependence 
of the in-medium effective nuclear interactions on heavy ion reactions 
at intermediate energies.
Apart the already known $Isoscalar$ effects, like less stopping, less 
compression and an overall
stiffer EoS, we have investigated in detail the $Isovector$ contributions, 
that give raise to
the poorly known neutron-proton effective mass splitting in asymmetric 
matter. We remind that this
point is also of interest for relativistic bosonic models of the nuclear 
matter 
\cite{fuchswci,baranPR,liu02,fuchsPPNP06} and the development of 
effective Lagranginas 
for non-perturbative QCD related to the breaking of the chiral symmetry
\cite{frank03,shao06,plu09}. We have shown that the fast nucleon and 
light ions emissions in
heavy ion collisions at intermediate energies are affected by the 
($n,p$) mass splitting.
The isospin yield ratios ($n/p, ^3H/^3He$) and elliptic flows appear 
more sensitive to 
the Mass-splitting than to the stiffness of the Symmetry Term at high 
density, in particular for
high transverse momentum selections. This opens the possibility of 
disentangling the two distinct
isospin effects on the reaction dynamics. New more accurate and exclusive 
data are needed. We
like to mention the new measurements that will be performed at SIS-GSI 
by the ASYEOS 
Collaboration \cite{lemmon09} and the new experiments planned at 
RIKEN-Tokyo and CSR-Lanzhou
also with unstable, more neutron-rich, beams.

\vskip 0.3cm
{\it Acknowledgements}

We warmly thank W.Reisdorf, W.Trautmann, P.Danielewicz, W.G.Lynch, Qinfeng Li 
and P.Russotto
for several nice and inspiring discussions.


\begin{thebibliography}{99}
\bibitem{equalmass}
We remark that, due to the different neutron/proton Fermi momentum
in isospin asymmetric matter, this parametrization is  different
from the one without 
$Isovector$ contributions 
to the momentum dependence, i.e. with the choice $z=C/8$, see 
Eq.(\ref{en_c}).
\end{thebibliography}
\end{document}